\documentclass[preprint,showpacs,preprintnumbers,amsmath,amssymb]{revtex4}

\usepackage{graphicx}
\usepackage{dcolumn}
\usepackage{bm}

\newcommand{\bq}{\begin{equation}}
\newcommand{\eq}{\end{equation}}
\newcommand{\bqa}{\begin{eqnarray}}
\newcommand{\eqa}{\end{eqnarray}}
\newcommand{\nn}{\nonumber \\}

\def\be     {\begin{equation}}
\def\ee     {\end{equation}}
\def\bea        {\begin{eqnarray}}
\def\eea        {\end{eqnarray}}
\def\bnn    {\begin{eqnarray*}}
\def\enn    {\end{eqnarray*}}

\begin{document}

\title{Sondheimer Oscillation as a Fingerprint of Surface Dirac Fermions}

\author{Heon-Jung Kim$^{1,\ast}$, Ki-Seok Kim$^{2,3}$, Mun Dae Kim$^{4}$, S.-J. Lee$^{5}$,
J.-W. Han$^{1}$, A. Ohnishi$^{5}$, M. Kitaura$^{5}$, and M.
Sasaki$^{5,\dagger}$,  A. Kondo$^{6}$, and K. Kindo$^{6}$
}
\affiliation{$^1$Department of Physics, College of
Natural Science, Daegu University, Gyeongbuk
712-714 Republic of Korea \\
$^2$Asia Pacific Center for Theoretical Physics, POSTECH, Pohang,
Gyeongbuk 790-784, Korea \\ $^3$Department of Physics, POSTECH, Pohang, Gyeongbuk 790-784, Korea \\
$^4$Institute of Physics and Applied Physics, Yonsei University,
Seoul 120-749, Korea \\ $^5$Department of Physics, Faculty of
Science, Yamagata University, Kojirakawa 1-4-12 Yamagata,
990-8560, Japan \\ $^6$Institute for Solid State Physics,
University of Tokyo, Kashiwanoha 5-1-5, Kashiwa, Chiba 277-8581
Japan}
\date{\today}

\begin{abstract}
Topological states of matter challenge the paradigm of symmetry
breaking,
%
%
characterized by gapless boundary modes and protected by the
topological property of the ground state.
%
%
Recently, angle-resolved photoemission spectroscopy (ARPES) has
revealed that semiconductors of Bi$_{2}$Se$_{3}$ and
Bi$_{2}$Te$_{3}$ belong to such a class of materials.
%
%
%
Here, we present undisputable evidence for the existence of
gapless surface Dirac fermions from transport in Bi$_{2}$Te$_{3}$.
We observe Sondheimer oscillation in magnetoresistance (MR).
%
%
This oscillation originates from the quantization of motion due to the confinement of
electrons within the surface layer. Based on Sondheimer's
transport theory, we determine the thickness of the surface state
from the oscillation data.
%
%
In addition, we uncover the topological nature of the surface state, fitting
consistently both the non-oscillatory part of MR and the Hall resistance.
 The side-jump contribution turns out to dominate
around $1$ T in Hall resistance while the Berry-curvature
effect dominates in $3$ T $\sim$ $4$ T.
\end{abstract}

\maketitle


Symmetry breaking is the paradigm in not only classifying quantum
states of matter but also describing phase transitions between
them, where the correlation length of fluctuations between local
order parameters diverges at the critical point of a continuous
transition \cite{Ginzburg,Nambu}. On the other hand, topological
states of matter are classified by topological quantum numbers
\cite{Hall1,Hall2}, associated with gapless boundary electronic
states and are protected from the topological properties of the ground
state \cite{TI_Review,FQHE_Review}. Instead of a divergence in
correlation length, topological phase transitions are accompanied by
changes of the gapless boundary modes. One possible mechanism for this phenomenon is
the length scale of the boundary state becomes of the same order as
the bulk size, causing the gapless modes in opposite boundaries  to be mixed
and making such boundary modes gapped \cite{Shindou}. In this
case, the length scale for the boundary mode plays basically the
same role as the correlation length for the phase transition,
which is the fundamental length scale for a topological phase
(Fig. 1a).

Recently, the semiconductor material Bi$_{2}$Se$_{3}$ and
Bi$_{2}$Te$_{3}$ have been verified to be three dimensional
topological insulators. These insulators are regarded as a novel quantum state of
matter \cite{TI_3D1,TI_3D2,TI_3D3-1,TI_3D3-2}, where gapless
surface electrons are uncovered in ARPES
\cite{Hasan_ARPES1,Hasan_ARPES2,ARPES_Shen}. Although the
surface states in these materials and the electronic structure in
graphene are described by Dirac theory (Fig. 1b), the surface
state of topological insulators is profoundly different from the
electronic structure of graphene \cite{Graphene_Review}. This difference
originates from the absence of both the sublattice symmetry and
 valley degeneracy.
%
%
The direction of spin is locked with that of the momentum in surface
Dirac electrons \cite{Hasan_ARPES3}. This completely suppresses
backscattering due to time-reversal invariant impurities,
 allowing a super-metallic state \cite{Ryu_TI}.
%
%

In this letter we focus on MR and Hall measurements, both of which
are of high importance for the fundamental understanding and practical
applications of topological insulators. We observe an oscillatory
behavior in MR at low magnetic fields of up to $4$ T. This behaviour can be
 identified as Sondheimer oscillation \cite{Sondheimer}, where the
oscillation period is linearly proportional to the magnetic field
(Fig. 2).
%
%
Sondheimer's transport theory \cite{Sondheimer}, combined with
 Dirac dispersion, enables us to determine the fundamental
length scale from our experimental data, which turns out to be
about $5$ atomic layers. In this respect the Sondheimer
oscillation can be regarded as an inevitable result of the surface
state.
%
%
The nature of the surface state can be explained by the single
Dirac-fermion theory. This explains not only the non-oscillatory part
of MR but also the topological properties implicit in Hall resistance
in a quantitative and consistent way (Fig. 4). In particular, we
show that the Hall resistance of the surface state is dominated by
the side-jump around 1 T and below while the Berry-curvature
contribution is dominant at higher fields (Fig. 5 in {\bf SI}
III).

In our measurements we used defect-controlled Bi$_2$Te$_3$ single
crystals. Usually, as-grown Bi$_2$Te$_3$ single crystals are
 p-doped because of the anti-site defects in Bi sites
\cite{Material}. In order to tune the Fermi level, we have
controlled the amount of defects by adding extra Bi or Te; the
doped Bi tends to increase the anti-site defects, while the doped
Te tends to decrease them. Based on this strategy, we have
succeeded in growing a range of Bi$_2$Te$_3$ single crystals, from
fully p-doped to fully n-doped regions. Though rare, we have
obtained almost insulating Bi$_2$Te$_3$ single crystals. The
chance for obtaining these samples is around 3 $\%$. The carrier
type is determined by thermoelectric power at room temperature and
also by Hall sign measured at 4.2 K. Samples $\sharp 1$ and $\sharp 2$
are hole-doped while sample $\sharp 5$ is
electron-doped. Samples $\sharp 3$ and $\sharp 4$ are close to
insulators, as such, they are expected to show the topological properties of the surface
state well. See the supplementary material ({\bf SI} I) for our
sample preparation.

Magnetoresistance (MR) and Hall effect measurements have been
carried out by a six-probe method at 4.2 K using a superconducting
magnet up to 4 T and a 60 T pulse magnet at ISSP in Tokyo
university up to 55 T. Here, the direction of the magnetic fields is
applied perpendicular to the naturally cleaved plane, on which the
current is applied. For the MR and Hall measurements, we
carefully contacted the lead wires reduce the induction noise
for high-field pulse-magnet experiments. We have
taken the anti-symmetrized and the symmetrized parts as Hall and
longitudinal resistances, respectively.

ARPES has unveiled only a single Dirac-fermion band at the surface
of Bi$_{2}$Te$_{3}$ \cite{Hasan_ARPES2,ARPES_Shen}. Therefore, in order to
analyze our experimental data, we introduce an
electromagnetic vector potential $\boldsymbol{\vec{ A}}$ and a
Zeeman term into the single Dirac-fermion theory. In addition, we
take into account impurity scattering at the level of a Born
approximation. Our theoretical analysis reveals that the
orbital contribution or the effect of the vector potential on
experimental data is irrelevant in the region of magnetic
fields below $4$ T ({\bf SI} IV). However, its influence on MR and Hall
resistance can appear at higher magnetic fields, where Landau levels
are fully developed.

We model the surface state as a thin layer with thickness $a$
(Fig. 1a), which is used in the Boltzmann equation ({\bf SI} II).
This approach is essentially the same as what Sondheimer used with
 metallic thin films except for the band structure, where
non-relativistic electrons are replaced with Dirac fermions in the
presence of the Zeeman term. The main consequence of this is the
quantization of motion along the direction normal to the surface which produces
 an oscillatory component of MR. In our measurements at magnetic fields below 4 T,
 the periodicity observed is
linearly proportional to $H$ \cite{Sondheimer}. This oscillation is
distinguished from Shubnikov-de Haas oscillation due to the
formation of Landau levels, where the periodicity is proportional
to $1/H$. See Fig. 2.
%
%

Sondheimer's transport theory \cite{Sondheimer} with the Dirac
dispersion gives the following expression for the oscillating part
of the longitudinal resistivity \bqa &&
\frac{\rho(H,T)}{\rho(0,0)} = \frac{1}{\kappa} \Re \phi(s) , \eqa
where the kernel \bqa && \frac{1}{\phi(s)} = \frac{1}{s} -
\frac{3}{8s^{2}} + \frac{3}{2s^{2}} \int_{1}^{\infty} d u e^{- s
u} \Bigl( \frac{1}{u^{3}} - \frac{1}{u^{5}} \Bigr) \eqa with
$s=\kappa+i\beta$ results from the distribution in the hard-wall
boundary condition for the $z$-direction ({\bf SI} II). Two
parameters appear in this expression, $\kappa = a/l$
and $\beta = a/r_{c}$, where $l$ is the mean free path and $r_{c}$
is the magnetic length (proportional to $1/H$). $a$ is the thickness
of the surface state. This transport theory produces the
$H$-linear periodicity. It is worth noting that the periodicity in
MR depends only on $\beta$ while $\kappa$ modifies the amplitude
of oscillation as shown in Fig. 3.

Our experimental data shows that the Sondheimer oscillation turns
into the Shubnikov-de Haas oscillation above $3$ T $\sim$ $4$ T
(Fig. 2). This is consistent with several recent transport measurements
that show Shubnikov-de Hass oscillation beginning at around $4$ T
\cite{Transport_SdH1,Transport_SdH2,Transport_SdH3,Transport_SdH4}.
The appearance of the Shubnikov-de Haas oscillation is the origin
for the mismatch of the oscillation amplitude.


We point out that $\beta$ can be written as $\beta = \frac{1}{2}
(k_{f} a) h (k_{f}/\bar{k})$ with the surface thickness $a$, where
$h=\hbar \omega_{L}/E_F$ is the ratio of the Zeeman energy $\hbar
\omega_{L}$ to the Fermi energy $E_{F}$ at the surface, and
$k_{f}$ and $\bar{k}$ are the Fermi wave-vector and the average
momentum, respectively. See the supplementary information for
details. An important aspect is that $h$ also enters both the
non-oscillatory part of MR and Hall resistance. Therefore, the
actual value of $h$ influences not only the periodicity of the
Sondheimer oscillation but also both the non-oscillatory part of
MR and Hall resistance. Precisely speaking, the surface thickness
cannot be determined from the Sondheimer oscillation alone.
Combined with the longitudinal and transverse resistances, we can
optimize the thickness $a$ and the coefficient $\gamma$
simultaneously ({\bf SI} II), where $\gamma$ is the ratio
$\hbar \omega_{L}/E_F$ at $H = 1$ T. It is interesting to note that the
optimized $\gamma$ almost coincides with the bulk value.
Furthermore, this $\gamma$ value seems to be common to both
Bi$_{2}$Te$_{3}$ and Bi$_{2}$Se$_{3}$ at a given field ({\bf SI}
II). Our fitting for the oscillatory part of MR, performed
consistently for both the non-oscillating part of MR and Hall
resistance, gives a result for  the surface thickness of approximately $5$
atomic layers. This is quite remarkable in that this value is
consistent with that in MBE-grown Bi$_2$Te$_3$ thin films
\cite{MBE_Thickness}.
%
%


Next, we focus on the topological nature of the surface state.
Strictly, the role of the single Dirac-fermion theory is not
essential in Sondheimer oscillation although Dirac
dispersion is utilized. It might be the case that the surface
state is realized due to the good surface quality of our samples.
However, we will show that the Sondheimer oscillation is a signature of
surface Dirac electrons in Bi$_{2}$Te$_{3}$, verifying that the
Hall resistance originates from the anomalous Hall effect of
Dirac theory. In addition, we show that the side-jump contribution dominates at low
magnetic fields and the Berry-curvature effect dominates at high magnetic
fields.
%
%

The single Dirac-fermion theory gives an analytic expression for
the longitudinal conductance \cite{Dirac_Hall_MR} \bqa &&
\sigma_{xx}(H,T \rightarrow 0) = \alpha \frac{e^{2}}{2 \pi \hbar}
\frac{\sqrt{1 + h^{2}}}{1 + 4 h^{2}} , \eqa where $\alpha \equiv
\frac{4 n e^{2} (2\pi \hbar)^{3} v_{f}}{m^{*} n_{I}
[V_{I}^{(0)}]^{2} k_{f}}$ is a dimensionless parameter which
measures the strength of disorder with an impurity density $n_{I}$
and an impurity potential $V_{I}^{(0)}$, while $v_{f}$ is the
Dirac-fermion velocity with the Fermi momentum $k_{f}$. $h$ is the
dimensionless magnetic field, introduced in Sondheimer
oscillation.

The same Dirac theory results in Hall conductance
\cite{Dirac_Hall_MR} \bqa \sigma_{xy}(H, T \rightarrow 0) &=&
\sigma_{xy}^{FS}(H, T \rightarrow 0) + \sigma_{xy}^{A}(H, T
\rightarrow 0) , \nn \sigma_{xy}^{A}(H, T \rightarrow 0) &=&
\sigma_{xy}^{B}(H, T \rightarrow 0) + \sigma_{xy}^{SJ}(H, T
\rightarrow 0) \nn &+& \sigma_{xy}^{SK}(H, T \rightarrow 0) . \eqa
Hall conductance consists of two contributions. The
first results from electrons near the Fermi surface, referred as
normal Hall conductance while the second contribution comes from a
fully occupied band, called anomalous Hall conductance. We use the
normal Hall conductance from the Boltzman equation approach for
the Sondheimer oscillation, which is given by \bqa \sigma_{xy}^{FS}(H, T
\rightarrow 0) = \kappa \sigma_{xx}(H, T \rightarrow 0) \frac{\Im
\phi(s)}{[\Re \phi(s)]^{2} + [\Im \phi(s)]^{2}} . \nonumber \eqa
The anomalous Hall conductance is also composed of two
contributions. The first comes purely from the topological
character of the band structure, identified with the
Berry-curvature term $\sigma_{xy}^{B}(H, T \rightarrow 0) = -
\frac{e^{2}}{4 \pi \hbar} \frac{h}{\sqrt{1 + h^{2}}}$, while the
second originates from scattering with disorder in the presence of
the spin-orbit interaction. This disorder contribution is
separated into the side-jump term $\sigma_{xy}^{SJ}(H, T
\rightarrow 0) = - \frac{e^{2}}{4 \pi \hbar} \frac{h}{\sqrt{1 +
h^{2}}} \Bigl\{ \frac{4}{1 + 4 h^{2}} + \frac{3}{(1 + 4
h^{2})^{2}} \Bigr\}$ and the skew scattering term
$\sigma_{xy}^{SK}(H, T \rightarrow 0) = - \eta \frac{e^{2}}{2 \pi
\hbar} \frac{h}{(1 + 4 h^{2})^{2}}$.
%
%
It is interesting to observe that the side-jump term does not
depend on the disorder strength.
The dimensionless parameter $\eta \equiv \frac{[V_{I}^{(1)}]^{3}
v_{f} k_{f}}{2 \pi n_{I} [V_{I}^{(0)}]^{4}}$ in the skew
scattering term measures the disorder strength in the third order,
where $V_{I}^{(1)}$ is a disorder strength of the third order.

Based on Eqs. (3) and (4), we obtain the longitudinal and Hall
resistances as follows \bqa && \rho_{xx}(H, T \rightarrow 0) =
\frac{\sigma_{xx}(H, T \rightarrow 0)}{[\sigma_{xx}(H, T
\rightarrow 0)]^{2} + [\sigma_{xy}(H, T \rightarrow 0)]^{2}} , \nn
&& \rho_{xy}(H, T \rightarrow 0) = \frac{\sigma_{xy}(H, T
\rightarrow 0)}{[\sigma_{xx}(H, T \rightarrow 0)]^{2} +
[\sigma_{xy}(H, T \rightarrow 0)]^{2}} . \eqa It should be noted
that $\sigma_{xy}$ in the denominator cannot be ignored in this
case because this term is comparable to $\sigma_{xx}$. Here we
have two dimensionless parameters,  $\alpha$ and $\eta$.
However, the contribution from the skew scattering turns out to be
negligible ({\bf SI} III). Only one fitting parameter $\alpha$ remains for
both MR and Hall resistance.




When either holes (sample $\sharp 1$ and $\sharp 2$) or electrons
(sample $\sharp 5$) are heavily doped, MR curves greatly deviate from the
single Dirac-fermion theory ({\bf SI} III). On the other
hand, nearly insulating samples (sample $\sharp 3$ and $\sharp 4$)
display reasonable matches between experiment and theory. see
Fig 4. The Hall resistance also shows deviation from the single
Dirac-fermion theory for heavily doped samples ({\bf SI} III) but
not much for nearly insulating samples (sample $\sharp 3$ and
$\sharp 4$), implying that insulating samples are explained by the
theory in a quantitative and consistent way. These results provide
a compelling evidence for surface Dirac electrons. In particular,
the dominant contribution in the Hall resistance turns out to be
the side-jump mechanism at fields below 1 T and the
Berry-curvature effect at higher fields. See Fig. 5 in {\bf SI} III.

It is also worth noting that the curvature of the Hall resistance
in sample $\sharp 4$ is larger than that in sample $\sharp 3$.
According to Dirac theory with disorder, two parameters affect
the shape of the Hall resistance: the disorder strength $\alpha$
and the parameter $\gamma$ that measures the distance from the
Dirac point. In Fig. 6 of {\bf SI} III, we show how these parameters
influence the curvature of Hall resistance. By decreasing $\gamma$,
the Hall resistance becomes straighter because the anomalous Hall
effect weakens.

In this letter we have measured the fundamental length scale of
the topological insulator, the thickness of the surface state,
from the Sondheimer oscillation in magnetoresistance. This surface
state is described by the single Dirac-fermion theory. The
topological nature of this is verified by the fact that the Hall
resistance mainly results from the anomalous Hall effect of Dirac
theory, which in turn is dominated by both the side-jump mechanism and the
Berry-curvature effect.
%
%

The surface thickness will diverge at the critical point of a
phase transition from a band insulator to a topological insulator.
Such a phase transition was demonstrated in the HgTe quantum well
structure when the size of the quantum well was tuned
\cite{Zhang_HgTe1,Zhang_HgTe2}. On the other hand, the topological
phase transition has not yet been achieved in three-dimensional
topological insulators. Our measurement for the surface thickness
can be utilized as an important tool, revealing the mechanism of
such a topological phase transition.


\section{Acknowledgements}

This research is supported by Basic Science Research Program
through the National Research Foundation of Korea (NRF) funded by
the Ministry of Education, Science, and Technology (No.
2010-0021438). K.-S. Kim is supported by the National Research
Foundation of Korea (NRF) grant funded by the Korea government
(MEST) (No. 2010-0074542).


$^{\ast}$hjkim76@daegu.ac.kr; $^{\dagger}$sasaki@sci.kj.yamagata-u.ac.jp \\

\begin{figure}[t]
\caption{{\bf A schematic picture of the surface layer with a
length scale $a$ and the Dirac cone at the surface state.} a. The
schematic diagram for Bi$_2$Te$_3$ shows the surface layer with
thickness $a$. The surface thickness $a$ is determined from
Sondheimer oscillation in magnetoresistance, originating from the
quantization of motion within the surface layer. b. The Dirac
dispersion of the surface state gives rise to topologically
nontrivial physical properties. In particular, the anomalous Hall
effect due to the Dirac cone turns out to dominate in the Hall
resistance. See Fig. 5 in the supplementary information.}
\label{fig1}
\end{figure}

\begin{figure} [t]
\caption{{\bf Sondheimer oscillation in magnetoresistance.} a. The
second derivative of the resistance with respect to the applied
magnetic field shows oscillation with a periodicity in $H$, compared
to the theoretical curve (red thick line) based on
Sondheimer's transport theory. The experimental periodicity
matches well with the theoretical curve, but it deviates from
theoretical values around $H$ = 4 T. b. Peak and dip number vs. peak
and dip position (magnetic fields) in (a). Data points are located on a straight
line at low magnetic fields, confirming the $H$ linear periodicity
instead of the $1/H$ periodicity. On the other hand, the
experimental data will deviate from the straight line in high
magnetic fields due to the appearance of Shubnikov-de Haas
oscillation. The second derivative of the magnetoresistance
measured up to 55 T is plotted with respect to $H$ (c) and 1/$H$
(d). This comparison reveals that the Sondheimer oscillation
exists in the region of low magnetic fields while the Shubnikov-de
Haas oscillation with a periodicity in 1/$H$ appears at high
magnetic fields. } \label{fig2}
\end{figure}

\begin{figure} [t]
\caption{{\bf Dependence of Sondheimer oscillation on the Fermi
energy and the disorder strength.} a. Dependence of Sondheimer
oscillation on the Fermi energy. $\gamma = h / H$ measures the
distance of the Fermi energy from the Dirac point, where $h =
\hbar \omega_{L} / E_{F}$ is the dimensionless magnetic field
given by the ratio of the Zeeman energy $\hbar \omega_{L}$ to the
Fermi energy $E_{F}$. Increasing $\gamma$, i.e., as the Fermi
surface becomes close to the Dirac point, the period of the
Sondheimer oscillation decreases. b. Dependence of Sondheimer
oscillation on the disorder strength. $\kappa=a/l$ measures the mean free path.
 It does not affect the periodicity, changing
the amplitude of the oscillation only. c. Peak and dip number vs. peak and dip position
 (magnetic fields) as a function of $\gamma$ with a fixed
$\kappa$. This confirms our conclusion in Fig. 3a. }\label{fig3}
\end{figure}

\begin{figure} [t]
\caption{{\bf Magnetoresistance and Hall resistance for sample
$\sharp 3$ and $\sharp 4$ with theoretical fitting.} The
magnetoresistance and Hall resistance of sample $\sharp 3$ are
displayed in a and b, respectively, together with theoretical
curves (red thick line) based on the single Dirac-fermion theory.
The same quantities of sample $\sharp 4$ are presented in c and d.
We emphasize that the theoretical curves for Hall data are based
on the parameters from our fitting of magnetoresistance data. In
other words, $\gamma$ (the ratio of the Zeeman energy to the
Fermi energy) and $\alpha$ (the disorder strength in the
longitudinal resistance) are determined from both the Sondheimer
oscillation and magnetoresistance completely, and there are no
free parameters for the Hall resistance. See the text. Reasonable
matches between the experimental data, particulary for the Hall
resistance, and the theory reveal that the nature of the surface
state is described by the single Dirac-fermion theory.
Furthermore, we found that the Berry-curvature term dominates the
experimental data of Hall resistance around $3$ T $\sim$ $4$ T while
the side-jump mechanism works around $H$ = $1$ T, confirming the
topological origin of the transport phenomena in the surface state
of topological insulators. See Fig. 5 in the supplementary
information. } \label{fig4}
\end{figure}

\begin{thebibliography}{9}

\bibitem{Ginzburg} V. L. Ginzburg, {\it Rev. Mod. Phys.} {\bf 76}, 981
(2004).


\bibitem{Nambu} Y. Nambu, {\it
Rev. Mod. Phys.} {\bf 81}, 1015 (2009).

\bibitem{Hall1} N. Nagaosa, J. Sinova, S. Onoda, S., A. H. MacDonald,
and N. P. Ong, {\it Rev. Mod. Phys.}
{\bf 82}, 1539 (2010).

\bibitem{Hall2} D. Xiao, M.-C. Chang, Q. Niu, {\it Rev. Mod.
Phys.} {\bf 82}, 1959 (2010).

\bibitem{TI_Review} M. Z. Hasan and C. L. Kane, {\it Rev. Mod. Phys.} {\bf 82}, 3045
(2010).

\bibitem{FQHE_Review} C. Nayak, S. H. Simon, A. Stern, M. Freedman, and S. D Sarma,
{\it Rev. Mod. Phys.} {\bf 80}, 1083 (2008).

\bibitem{Shindou} R. Shindou, R. Nakai, and S. Murakam, {\it New J. of Phys.} {\bf 12}, 065008
(2010).

\bibitem{TI_3D1} L. Fu, C. L. Kane, and E. J. Mele, {\it Phys. Rev. Lett.}
{\bf 98}, 106803 (2007).

\bibitem{TI_3D2} J. E. Moore and L. Balents, {\it Phys.
Rev. B} {\bf 75}, 121306 (2007).

\bibitem{TI_3D3-1} R. Roy, {\it Phys. Rev. B} {\bf 79},
195322 (2009).

\bibitem{TI_3D3-2} R. Roy, {\it Phys. Rev. B} {\bf 79}, 195321 (2009). 


\bibitem{Hasan_ARPES1} D. Hsieh, et al., {\it Nature} {\bf 452},
970 (2008).

\bibitem{Hasan_ARPES2} D. Hsieh et al., {\it Science} {\bf 323}, 919 (2009).

\bibitem{ARPES_Shen} Y. L. Chen et al., {\it
Science} {\bf 10}, 178 (2009).

\bibitem{Graphene_Review} A. H. Castro Neto et al., {\it Rev. Mod. Phys.} {\bf 81}, 109
(2009).

\bibitem{Hasan_ARPES3} D. Hsieh et al., {\it Nature}
{\bf 460}, 1101 (2009).

\bibitem{Ryu_TI} K. Nomura, M. Koshino, and S. Ryu, {\it
Phys. Rev. Lett.} {\bf 99}, 146806 (2007).

\bibitem{Sondheimer} E. H. Sondheimer, {\it Phys. Rev.} {\bf 80}, 401 (1950).

\bibitem{Transport_SdH1} A. A. Taskin and Y. Ando, {\it Phys. Rev. B} {\bf 80}, 085303 (2009).

\bibitem{Transport_SdH2} D. Qu et al., {\it Science} {\bf 329}, 821 (2010).

\bibitem{Transport_SdH3} J. G. Analytis et al., {\it Nature Physics} {\bf 6}, 960 (2010). 


\bibitem{Transport_SdH4} Z. Ren, et al., {\it Phys. Rev. B} {\bf  82}, 241306 (2010).

\bibitem{MBE_Thickness} C.-L. Song, Preprint at http://arxiv.org/abs/1007.0809 (2010).

\bibitem{Dirac_Hall_MR} N. A. Sinitsyn et al., {\it Phys. Rev. B} {\bf 75}, 045315 (2007).


\bibitem{Zhang_HgTe1} B. A. Bernevig, T. L. Hughes,
and S. C. Zhang, {\it Science} {\bf 15},
1757 (2006).

\bibitem{Zhang_HgTe2} M. Konig et al., {\it Science} {\bf 2}, 766 (2007).

\bibitem{Material} Y. S. Hor et al., {\it Phys. Rev.} B {\bf 79}, 195208 (2009).



\end{thebibliography}
\end{document}


\title{Supplementary information for Sondheimer oscillation as a fingerprint of surface Dirac fermions}
\author{Heon-Jung Kim$^{1,\ast}$, Ki-Seok Kim$^{2,3}$, Mun Dae Kim$^{4}$, S.-J. Lee$^{5}$,
J.-W. Han$^{1}$, A. Ohnishi$^{5}$, M. Kitaura$^{5}$, and M.
Sasaki$^{5,\dagger}$,  A. Kondo$^{6}$, and K. Kindo$^{6}$ }
\affiliation{$^1$Department of Physics, College of Natural
Science, Daegu University, Gyeongbuk
712-714 Republic of Korea \\
$^2$Asia Pacific Center for Theoretical Physics, POSTECH, Pohang,
Gyeongbuk 790-784, Korea \\ $^3$Department of Physics, POSTECH, Pohang, Gyeongbuk 790-784, Korea \\
$^4$Institute of Physics and Applied Physics, Yonsei University,
Seoul 120-749, Korea \\ $^5$Department of Physics, Faculty of
Science, Yamagata University, Kojirakawa 1-4-12 Yamagata,
990-8560, Japan \\ $^6$Institute for Solid State Physics,
University of Tokyo, Kashiwanoha 5-1-5, Kashiwa, Chiba 277-8581
Japan}
\date{\today}

\begin{abstract}
In this supplementary information we discuss how to fit our
experimental data to theoretical predictions of transport on the surface of Bi$_2$Te$_3$.
First, we describe the sample synthesis and its resulting character.
Second, we present the detailed
procedure for fitting Sondheimer oscillation data to theory, where the fundamental
length scale of the surface state is determined. Third, we discuss
the non-oscillatory part of magnetoresistance and Hall resistance
based on the single Dirac-fermion theory, where the topological
nature of the surface state is verified. Last, we consider the
quantum Hall effect by examining electromagnetic vector
potential. The theoretical result from the quantum Hall effect shows
that our experimental regime is far from
the quantum Hall effect and thus, the effect of the electromagnetic vector
potential is negligible in our analysis.
\end{abstract}

\maketitle

\section{Sample preparation and characterization}

Single crystals of Bi$_2$Te$_3$ were grown by a modified
bridgeman method, where Bi$_2$Te$_3$ powder is melted and
crystalized in an evacuated quartz ampoule several times by slow
cooling. The sample was cooled from  850$^{\circ}$ C to
550$^{\circ}$ C with a cooling rate of -10 K/h. Since as-grown
Bi$_2$Te$_3$ is usually p-doped (p-Bi$_2$Te$_3$) due to the
anti-site defects at the Bi sites\cite{Material}, we have
controlled the amounts of these defects  by adding extra Bi or Te;
the doped Bi tends to increase such defects, while the doped Te
tends to reduce them. To grow topologically insulating samples
(TI-Bi$_2$Te$_3$), we should add small amounts of Te to the
p-Bi$_2$Te$_3$ ingots. In our experiments, we found that $7 \sim
8$ $\%$ extra Te was necessary in order to synthesize
TI-Bi$_2$Te$_3$. Even if such an ingot were carefully grown,  only
few single crystals from that ingot showed the desired nonmetallic T-dependent
resistivity curves, indicating that large ingots usually contain
weak  concentration- or defect-gradient \cite{ong10}. Therefore,
we had to select and check every crystal from the large ingots to
confirm their non-metallic characteristics. In the selection process, we
measured the temperature-dependent resistivity down to 77 K and
thermoelectric power at room temperature of about 200 samples. The
n-Bi$_2$Te$_3$ single crystals were obtained by the further doping of
Te to produce the TI-Bi$_2$Te$_3$ ingots.

For our experiments, we selected two p-Bi$_2$Te$_3$ samples (fully p-type
$\sharp 1$ and lightly p-type $\sharp 2$), two TI-Bi$_2$Te$_3$ samples
($\sharp 3$ and $\sharp 4$), and one n-Bi$_2$Te$_3$ sample ($\sharp 5$).
The carrier sign was determined from
the sign of the thermopower at room temperature and was also checked using
Hall resistivity at 4.2 K, both of which gave the same sign.
Magnetoresistance (MR) and Hall effect measurements have been
carried out by a six-probe method at 4.2 K, using a
superconducting magnet up to 4 T and a 60 T pulse magnet at ISSP
in the university of Tokyo up to 55 T. Here, the direction of the magnetic field
 applied is perpendicular to the naturally cleaved plane, on
which the current is applied. For MR and Hall measurements, we
 carefully contacted the lead wires to reduce induction noise 
 for high-field pulse-magnet experiments. We
have taken the anti-symmetrized and symmetrized parts in Hall and
longitudinal resistances, respectively.

Figure 1 shows the temperature dependency of the resistivity for the p-,
TI-, and n-Bi$_2$Te$_3$ single crystals. The resistivity for the both
p- and n-type samples ($\sharp 1$, $\sharp 2$, and $\sharp 5$)
decreases monotonically with decreasing temperature, which is a
typical metallic characteristic, while the resistivity for the TI
samples ($\sharp 3$ and $\sharp 4$) increases below $\sim$180 K
and then tends to saturate below 50 K. The magnitude of the
resistivity for the TI-Bi$_2$Te$_3$ single crystals is $5 \sim 10$
times larger than those for the p- and n-type samples. The
non-metallic nature observed for the TI samples is consistent with
those reported by the Princeton group \cite{ong10}. The oscillating
part of MR was separated by fitting the background with a
polynomial.

\begin{figure}[t]
\includegraphics[width=0.5\textwidth]{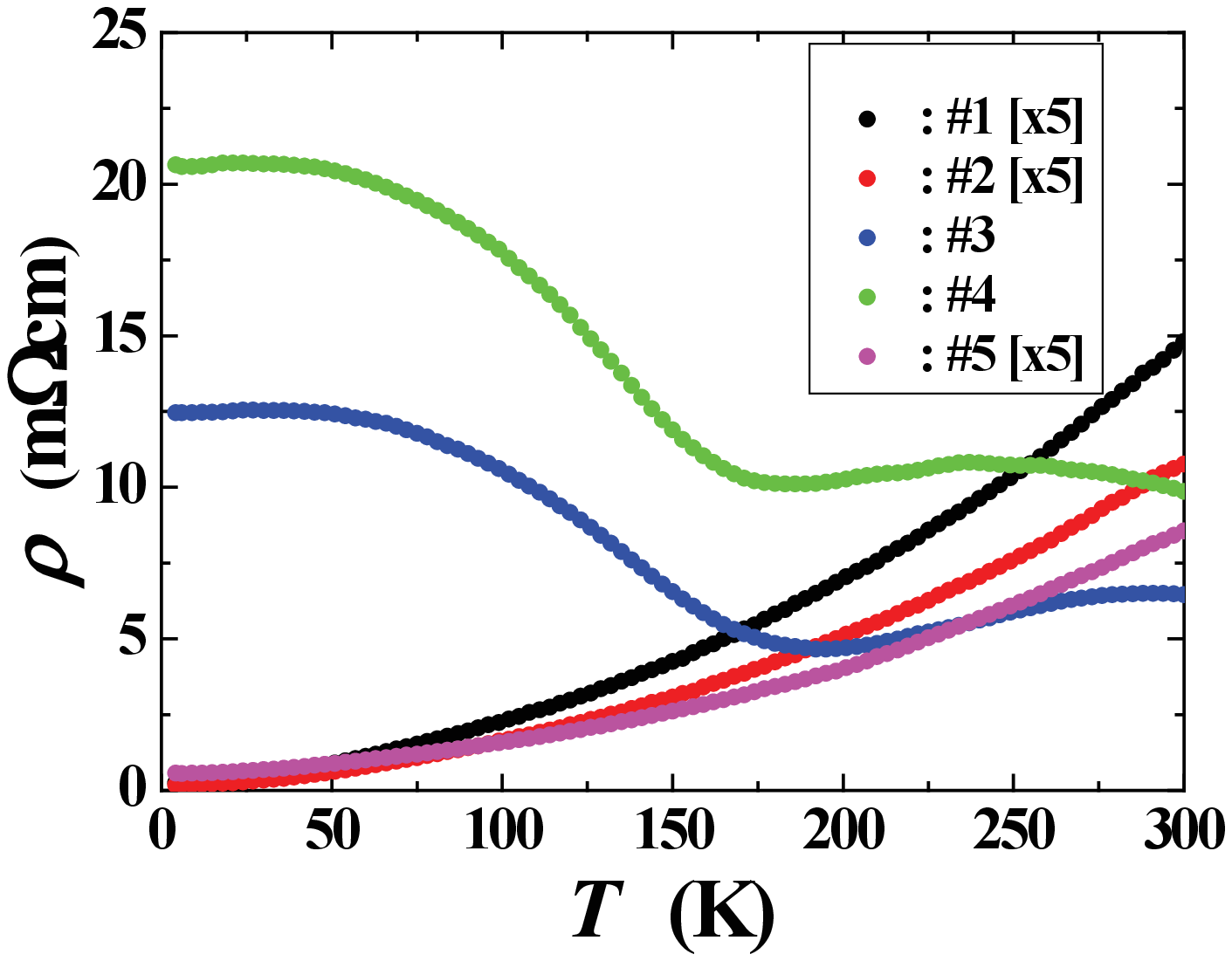}
\caption{Temperature dependency of resistivity for samples
$\sharp 1$ - $\sharp 5$. } \label{fig1}
\end{figure}

We have analyzed all five samples and discussed them in both the main text and
the supplementary information. Among them, the TI samples ($\sharp
3$ and $\sharp 4$) are best fitted to the single
Dirac-fermion theory while the fitting becomes worse for the p- and
n-Bi$_2$Te$_3$ samples. Deviation between the data of samples
$\sharp 1$, $\sharp 2$, and $\sharp 5$ and the theoretical curves can
be seen in Fig. 3 and Fig. 4. These figures clearly show that the
linear part in MR of the experimental data deviates from the
theoretical fitting significantly, implying that bulk channels for
conduction are important beyond the single Dirac-fermion theory.
Judging from our analysis, the Fermi level of the surface states
for sample $\sharp 3$ and $\sharp 4$ lies below the Dirac point,
corresponding to hole doping, while that in the bulk is between
the bulk gap. The difference of the Fermi level between the
surface and bulk originates from a band bending as reported before \cite{Transport_SdH3,Bi2Te3}.
On the other hand, the bulk Fermi level is below the valence band
maximum for sample $\sharp 1$ and $\sharp 2$. This explains the
metallic properties and the disagreement between theory and
experimental data for samples $\sharp 1$, $\sharp 2$, and $\sharp 5$.

\section{Sondheimer oscillation}

 Sondheimer oscillation  appears in metallic thin
films, where a finite thickness of a surface state gives rise to a
quantization of motion, resulting in an oscillation in
MR and Hall resistance \cite{Sondheimer}.
While Shubnikov-de Haas oscillation displays periodicity
proportional to $1/H$, the periodicty of Sondheimer oscillation is proportional to
 $H$. This periodicity allows us to find the
thickness of the surface state. Although Bi$_{2}$Te$_{3}$ is a
bulk material (semiconductor), the topological structure of the
ground state gives rise to a surface state with thickness $a$, which is
protected from time-reversal invariant perturbations
\cite{TI_3D1,TI_3D2,TI_3D3-1,TI_3D3-2}. In this respect, the Sondheimer oscillation is
undisputable evidence for existence of a  surface state.

We start from the Boltzmann equation \bqa && - \frac{e}{\hbar}
\Bigl( \boldsymbol{E} + \frac{\boldsymbol{\bar{v}}}{c} \times
\boldsymbol{H} \Bigr) \cdot \boldsymbol{\nabla}_{\boldsymbol{k}} f
+ \boldsymbol{\bar{v}} \cdot \boldsymbol{\nabla}_{\boldsymbol{r}}
f = - \frac{f - f_{0}}{\tau} . \eqa $f = f_{0} +
f_{1}(\boldsymbol{\bar{v}},z)$ is a non-equilibrium distribution
function with its equilibrium part $f_{0}$. The non-equilibrium
part $f_1$ depends on the $z$ coordinate. $\boldsymbol{\bar{v}} = \hbar
\boldsymbol{\bar{k}}/m^{*}$ is the average velocity, where $m^{*}$
is an effective mass of the surface Dirac electrons, and
$\boldsymbol{\bar{k}}$ is the average momentum, which is determined later.
The dispersion is given by $\epsilon_{\boldsymbol{k}} = - \mu +
\sqrt{v_{f}^{2} k^{2} + (g^{*} H)^{2}}$ in the presence of
a z-directional magnetic field $H$, where $v_{f}$ is the Dirac
velocity and the Fermi momentum is
$k_{f} = \frac{\sqrt{\mu^{2} - (g^{*} H)^{2}}}{v_{f}}$. $\mu$ is
the chemical potential at the surface, and $g^{*}$ is the Lande-g
factor of the surface electron. $\boldsymbol{H} = H \hat{z}$ is an
applied magnetic field in the $z$ direction and $\boldsymbol{E} =
E_{x} \hat{x} + E_{y} \hat{y}$ is an electric field, where a
$y$-directional electric field is induced. $\tau$ is the mean-free
time, which measures the strength of disorder.

Following the same procedure as in the original paper of
Sondheimer \cite{Sondheimer}, we obtain \bqa && \rho(H,T) =
\frac{\rho_{0}}{\kappa} \Re \phi(s) ,
\eqa where $\rho(H,T)$ is the resistivity. $\rho_{0} =
\frac{m^{*}}{n e^{2} \tau}$ is the residual resistivity with the
density $n = \frac{8 \pi}{3} \Bigl(\frac{m^{*} \bar{v}}{2 \pi
\hbar}\Bigr)^{3}$ of surface electrons, and $\kappa =
\frac{a}{l}$, where $l = \bar{v} \tau$ is the mean free path and
$a$ is the surface thickness, determined from fitting. $\phi(s)$
results from the distribution function in the hard-wall boundary
condition for the $z$ direction, given by \bqa &&
\frac{1}{\phi(s)} = \frac{1}{s} - \frac{3}{8s^{2}} +
\frac{3}{2s^{2}} \int_{1}^{\infty} d u e^{- s u} \Bigl(
\frac{1}{u^{3}} - \frac{1}{u^{5}} \Bigr) \eqa after integrating
over $z$. $s = \kappa + i \beta$, where $\beta = \frac{a}{r_{c}}$
with the magnetic length $r_{c} = \frac{m^{*} \bar{v} c}{e H}$.

For the numerical analysis, it is important to express the $\beta$
variable in terms of dimensionless parameters because the surface
thickness $a$ is determined from an appropriate choice of one
parameter in $\beta$, referred to as $\gamma$, which will be
discussed later. It is given by \bqa && \beta = \frac{1}{2} (k_{f}
a) \frac{\hbar \omega_{L}}{E_{F}} \frac{k_{f}}{\bar{k}} , \eqa
where $\hbar \omega_{L} = \hbar \frac{e g^{*} H}{2m^{*} c}$ is the
effective Zeeman energy and $E_{F} = \frac{k_{f}^{2}}{2m^{*}}$ is
the Fermi energy with an effective mass $m^{*}$. $\bar{k}/k_{f}$
is determined from \bqa && \frac{\bar{k}}{k_{f}} = \frac{1}{t}
\int_{0}^{\infty} d \epsilon \frac{\epsilon}{\sqrt{\epsilon^{2} +
h^{2}}} \frac{e^{\frac{\sqrt{\epsilon^{2} +
h^{2}}-\mu^{\prime}}{t}}}{(e^{\frac{\sqrt{\epsilon^{2} +
h^{2}}-\mu^{\prime}}{t}} + 1)^{2}} , \eqa where we introduce
several dimensionless parameters, scaled by the Fermi energy, such
that an effective dispersion $\epsilon = v k / E_{F}$ with $v =
\frac{\hbar^{2} k_{f}}{2m^{*}}$, the Zeeman energy $h = \hbar
\omega_{L} / E_{F}$, the chemical potential $\mu^{\prime} =
\mu/E_{F} = \sqrt{1 + h^{2}}$, and an effective temperature $t = T
/ E_{F}$.

Based on this formulation, we fit the oscillation data of MR. See
Fig. 2 for sample $\sharp 2$. First, we determine $\kappa \approx
0.02$, which characterizes the strength of the disorder for the best match
of the amplitude with the Sondheimer oscillation.
It is of note the matching cannot be perfect. In fact, the variation of $\kappa$ does not
modify the periodicity of the Sondheimer oscillation although it
changes the oscillation amplitude.

The surface thickness is determined from the periodicity $\Delta
H$ in the experimental data.
Considering that Eq. (5) fixes the ratio of $\bar{k}/k_{f}$, two
parameters in $\beta$ remain unknown, which correspond to
$k_{f} a$ and $\gamma$ which is the ratio $\hbar \omega/E_F$ at $H$ = 1 T.
Therefore,
 we cannot determine the thickness of the surface state
from the oscillation period alone. As discussed in the main text,
the same value of $\gamma$ is utilized for both the
non-oscillatory part of MR and Hall resistance. If the value of
$\gamma$ were incorrectly chosen, we could not fit both the
non-oscillatory part of MR and Hall resistance consistently. See
Fig. 6-b. Combined with both longitudinal and transverse
resistances, we found that $\gamma \approx 0.44$ fits the experimental data well. This value is
quite interesting because it almost coincides with the $\gamma$
value determined from the values of the bulk parameters. For
example, if we use bulk values of the Lande-$g$ factor and
effective mass, which are  $g^{*} \approx 13.7$ and $m^{*} \approx
0.1 m_{e}$ ($m_{e}$ is the bare mass of an electron),
we obtain $\hbar \omega_{L} \approx 100$ K at $H =
1$ T, giving rise to $h \approx 0.44$ with a typical bulk value
for the Fermi energy $E_{F} \approx 230$ K. Since we cannot
determine $g^{*}$, $m^{*}$, and $E_{F}$ at the surface from our
experiment, each value at the surface may not be the same as the
bulk one. However, it should be noted that our $\gamma$ seems to
be universal in both Bi$_{2}$Te$_{3}$ and Bi$_{2}$Se$_{3}$
although both $g^{*}$ and $E_{F}$ in Bi$_{2}$Se$_{3}$ are almost
four times larger than those in Bi$_{2}$Te$_{3}$
\cite{Transport_SdH3}.

\begin{figure}[t]
\includegraphics[width=0.4\textwidth]{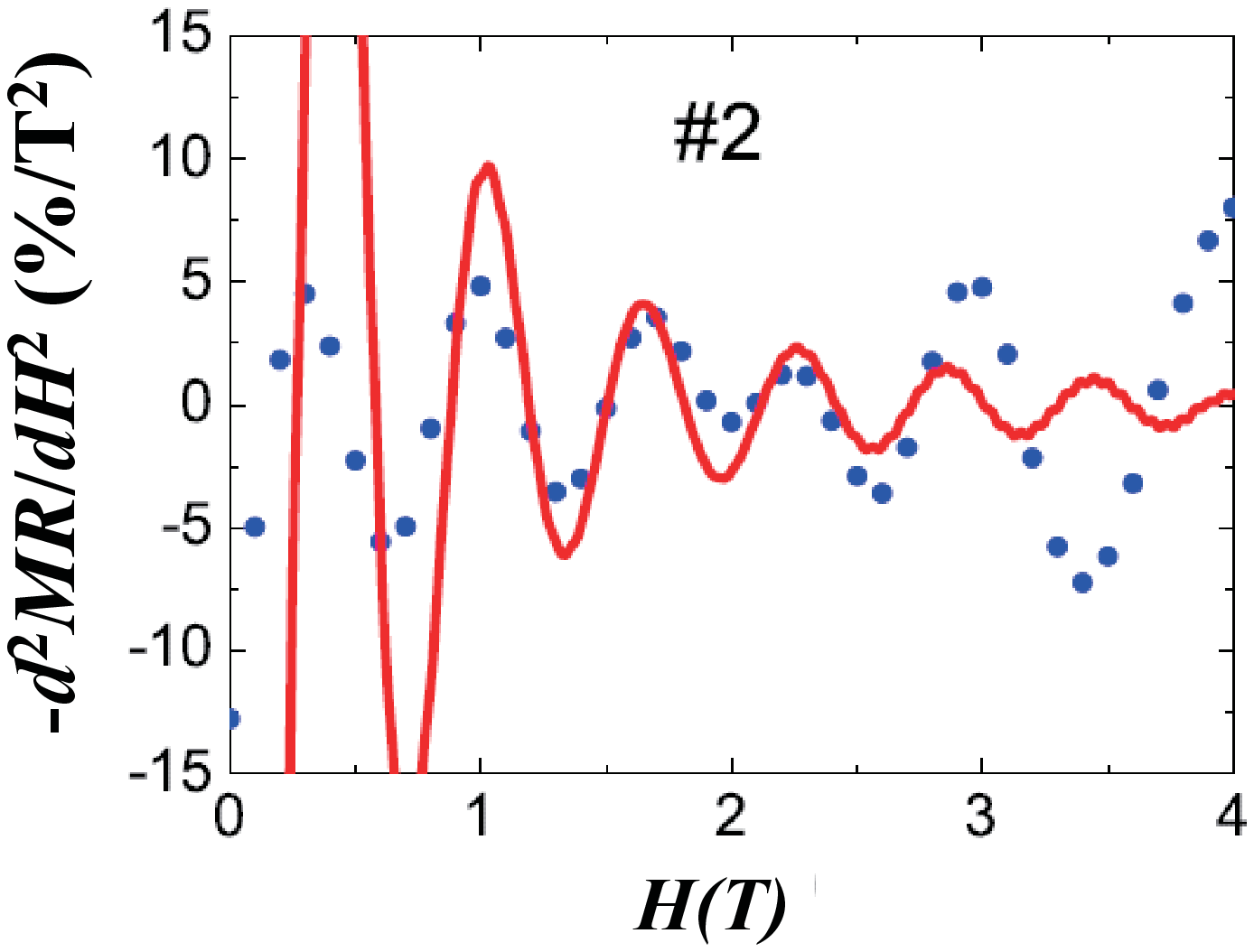}
\includegraphics[width=0.4\textwidth]{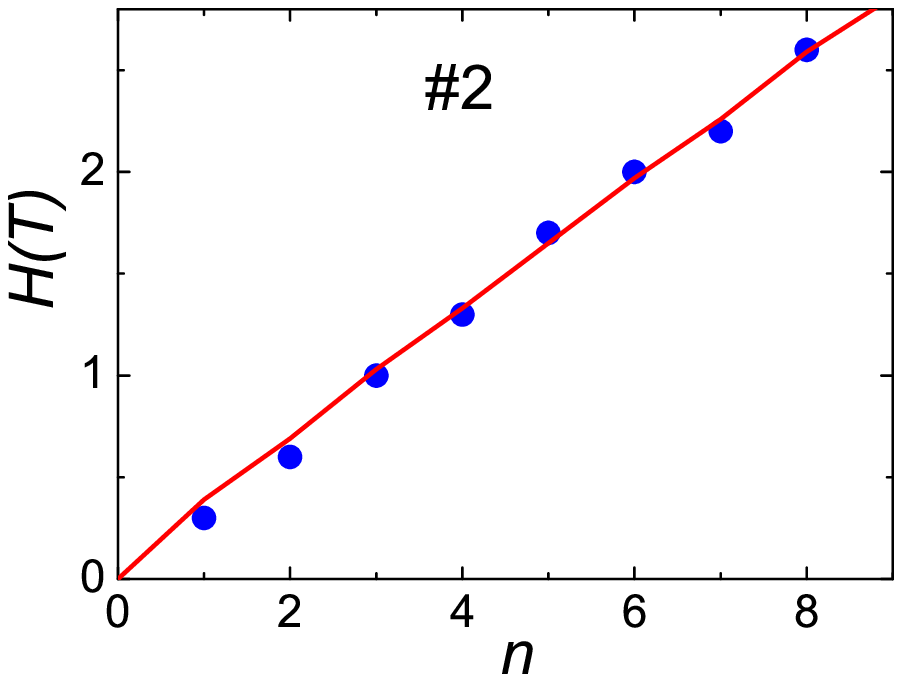}
\caption{a. Sondheimer oscillation for sample $\sharp 2$. The
periodicity is well matched, but the oscillation amplitude
deviates rather a lot due to the appearance of the Shubnikov-de
Haas oscillation above $4$ T. Sondheimer oscillations are also
observed for the p- and n-doped samples, where the surface channel for
conduction coexists with the bulk conduction. b. Peak and dip number vs. peak and dip position
 (magnetic fields) in (a). Data points are located on a
straight line, confirming the $H$ linear periodicity instead of
the $1/H$ periodicity.} \label{fig2}
\end{figure}

\section{Magnetoresistance and Hall effect}

Theoretically, the longitudinal resistance results from the transport of electrons
near the Fermi surface, and  is described quasi-classically or quantum-mechanically.
In this case,  the quasi-classical treatment based on
the Boltzmann equation gives the same result as the quantum
mechanical treatment based on the Kubo-formula. On the other hand,
there are various contributions with a topological origin in the
Hall resistance, which is beyond the conventional treatment used in the
Boltzmann equation approach. The Boltzmann equation needs additional
terms \cite{Dirac_Hall_MR} in order to mimic
the Kubo-formula \cite{Hall1,Hall2}.

\begin{figure}[t]
\includegraphics[width=0.4\textwidth]{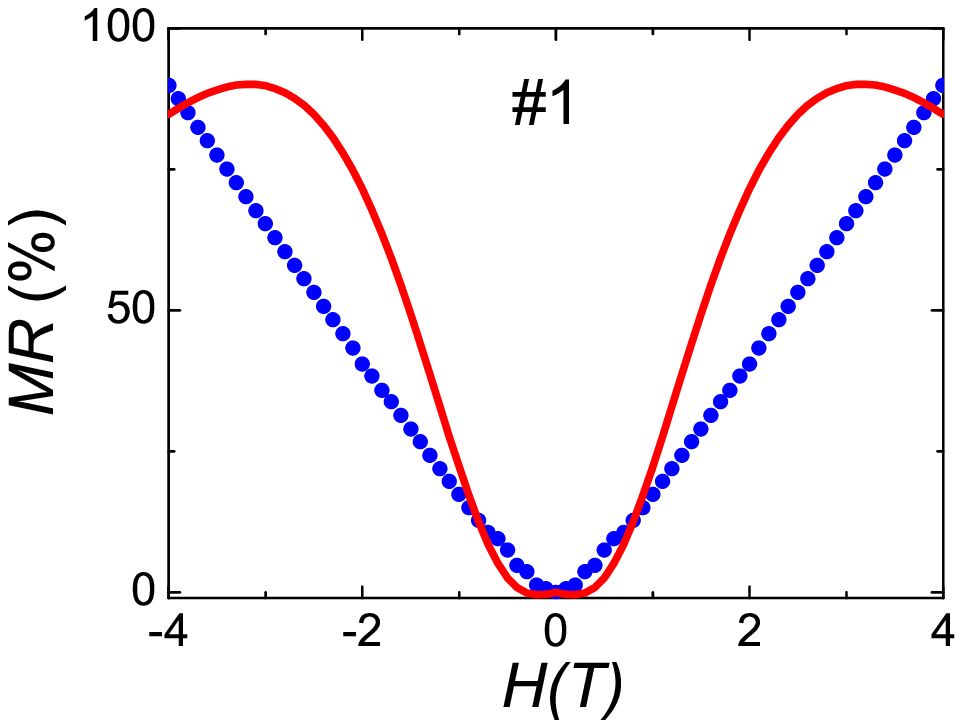}
\includegraphics[width=0.4\textwidth]{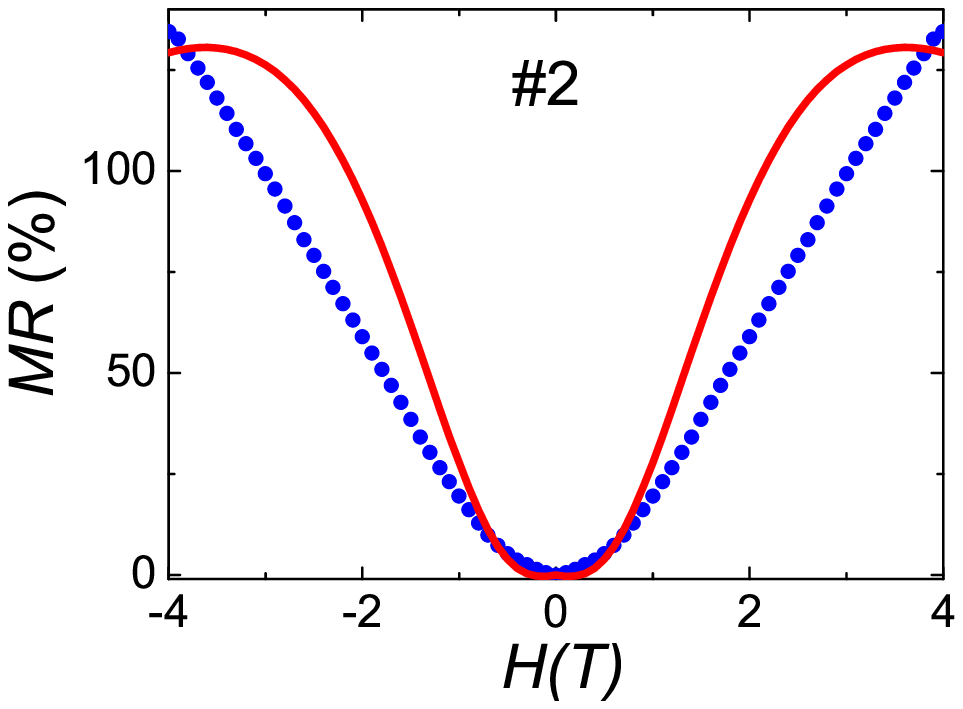}
\includegraphics[width=0.4\textwidth]{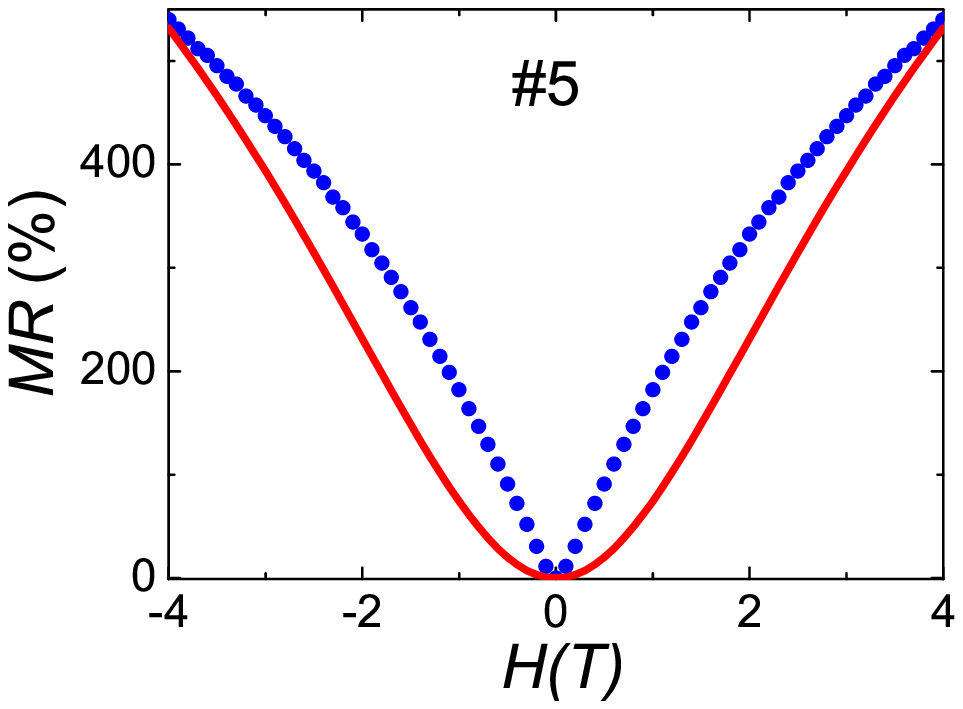}
\caption{Magnetoresistance for a. $\sharp 1$, b. $\sharp 2$, and
c. $\sharp 5$, where the disorder strength is utilized as a
fitting parameter, given by $\alpha \approx 0.4$, $\alpha \approx
0.5$, and $\alpha \approx 1.5$, respectively. As discussed in
section I, the presence of bulk conduction channels does not allow us
to describe these samples purely within the single Dirac-fermion theory.}
\label{fig3}
\end{figure}

\begin{figure}[t]
\includegraphics[width=0.4\textwidth]{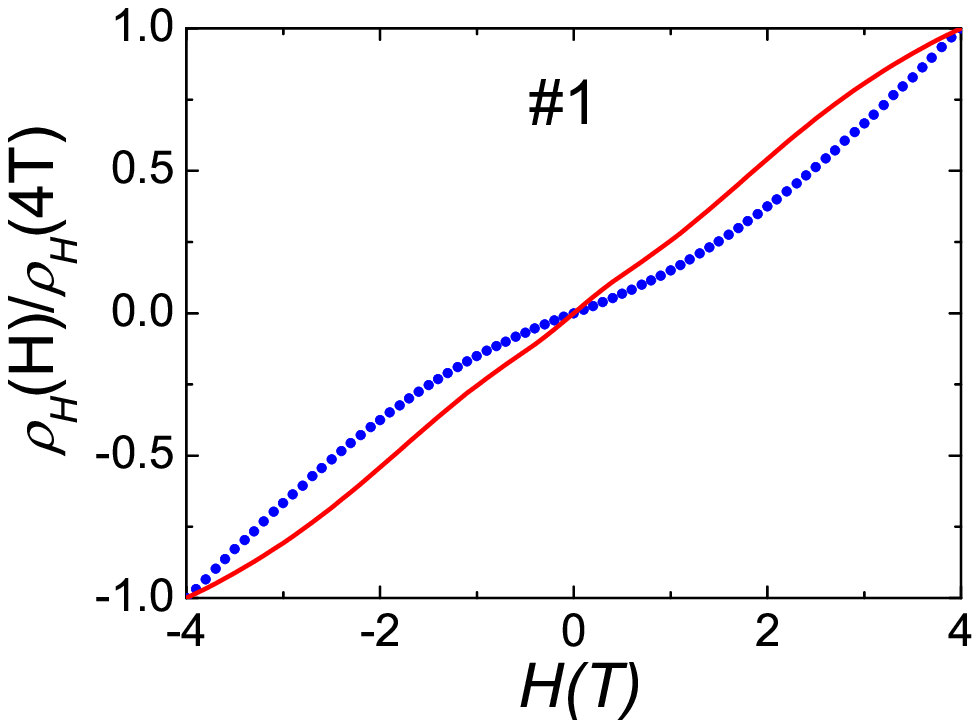}
\includegraphics[width=0.4\textwidth]{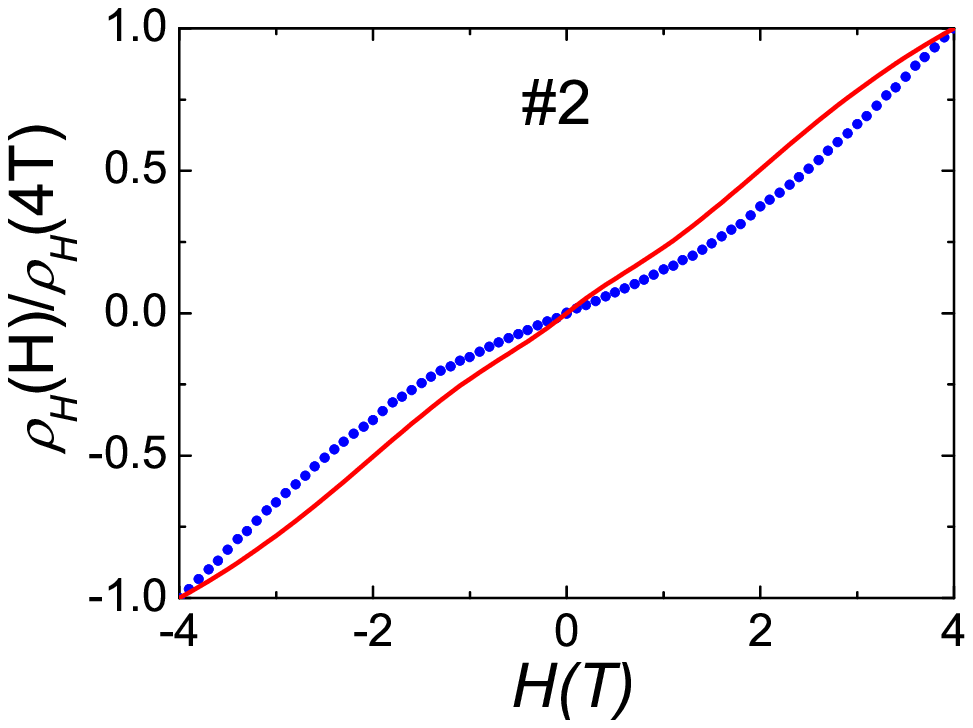}
\includegraphics[width=0.4\textwidth]{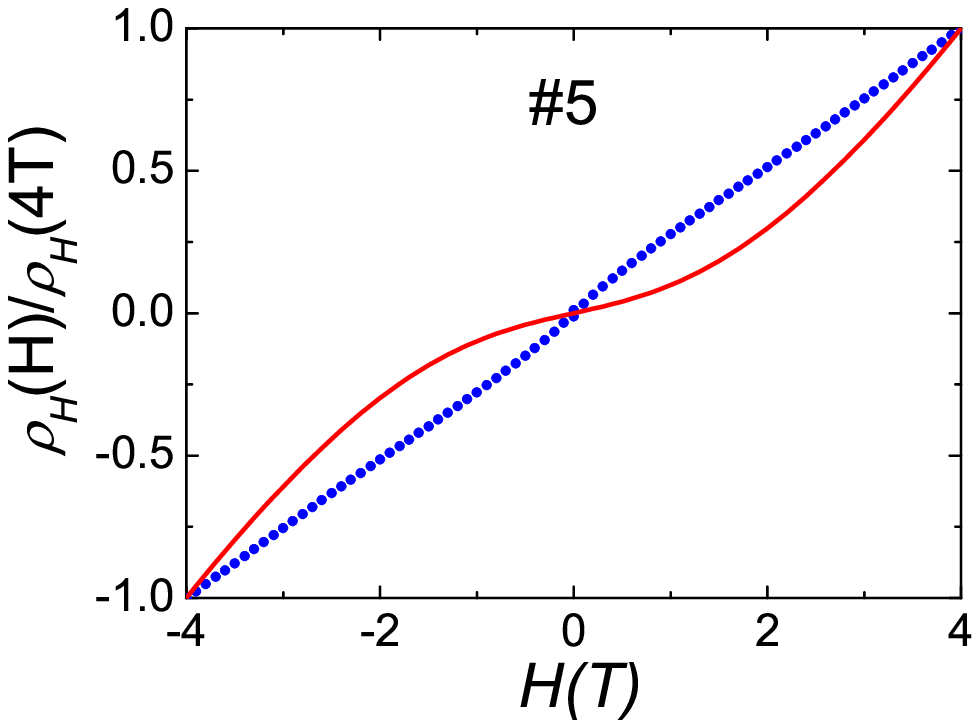}
\caption{Hall resistance for a. $\sharp 1$, b. $\sharp 2$, and c
$\sharp 5$, where the same strength of disorder is utilized as in
Fig. 3, respectively. We believe that the origin of this mismatch
between experiment and theory lies in the contribution from
the bulk transport which is not included in the single Dirac-fermion theory.}
\label{fig4}
\end{figure}



Since several terms exist in the Hall resistance, which are explained in
the main text, an important question is which terms give a large
contribution to the Hall resistance. We explicitly demonstrate that
the side-jump mechanism dominates in low magnetic fields while the
Berry-curvature effect dominates the Hall resistance at high
magnetic fields (Fig. 5). In contrast, the skew
scattering term turns out to be negligible. We also check the
dependence of the Hall resistance on either the disorder strength
$\alpha$ (Fig. 6-a) or the parameter $\gamma$, which determines
the Fermi level (Fig. 6-b). Roughly speaking, to vary $\gamma$ is to rescale the
$x$-axis while to change $\alpha$ is to rescale the $y$-axis.
 Figure 6-a and Fig. 6-b show how these parameters
affect the shape of the Hall resistance. These results confirm that if we do
not choose the correct $\gamma$, it is difficult to get a
reasonable match between the experiment and theory.

\begin{figure}[t]
\includegraphics[width=0.4\textwidth]{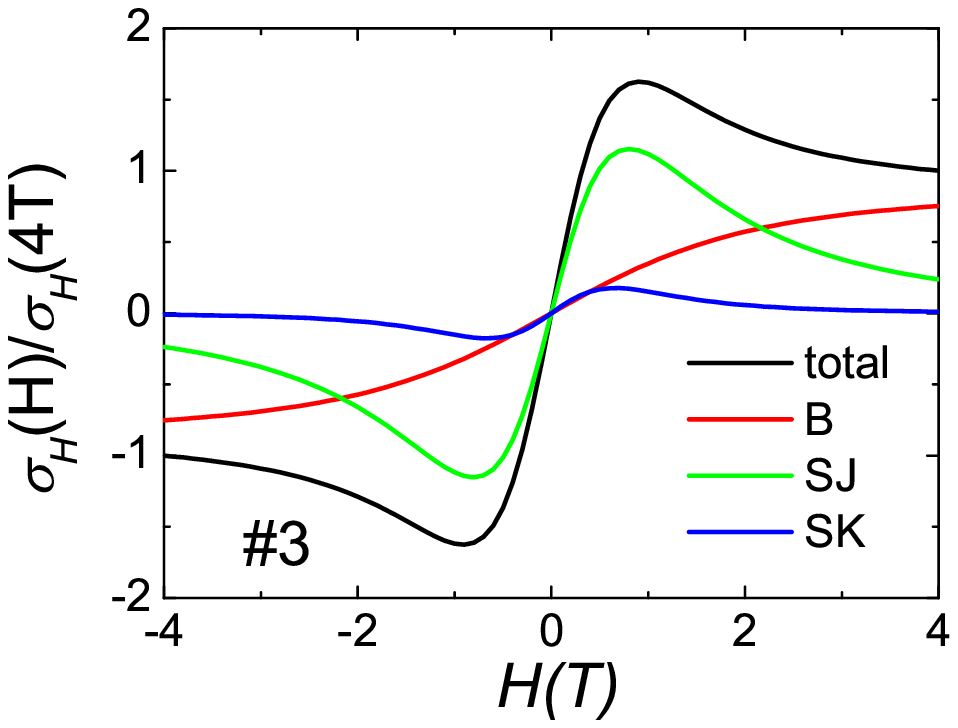}
\caption{Each contribution in Hall conductance shows that the
dominant term is the side-jump term in low fields and the Berry
curvature term in high fields.
 These curves are generated, by using the parameter values from sample $\sharp 3$,
  $\alpha = 0.5$ and $\gamma = 0.44$.}
  \label{fig7}
\end{figure}

\begin{figure}[t]
\includegraphics[width=0.4\textwidth]{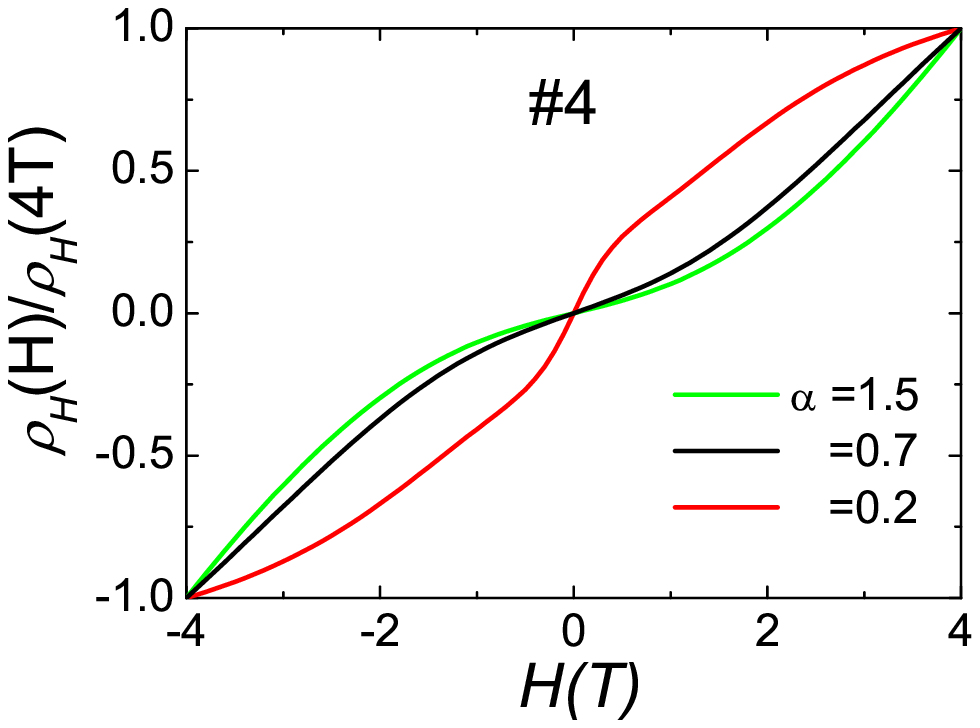}
\includegraphics[width=0.4\textwidth]{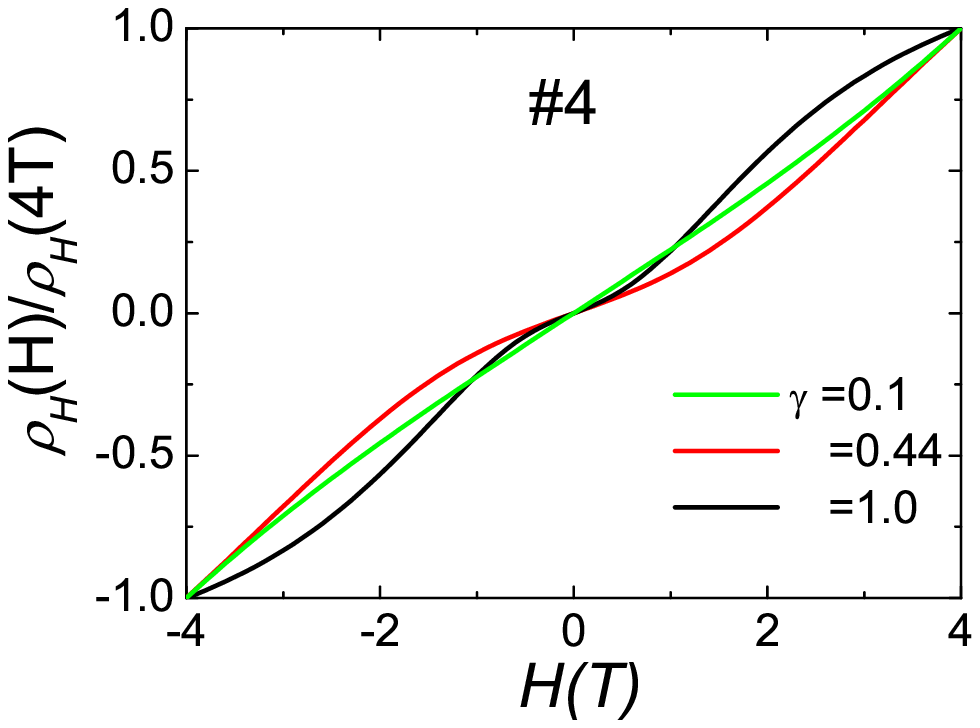}
\caption{a. Hall resistance deviates from the $\alpha = 0.7$ line
with $\gamma = 0.44$, changing $\alpha$. b. Hall resistance
separates from the $\gamma = 0.44$ line with $\alpha = 0.7$,
varying $\gamma$. The Hall resistance of sample $\sharp 4$ is well
fitted by using $\alpha = 0.7$ and $\gamma = 0.44$. } \label{fig8}
\end{figure}

\section{Contribution from Landau level}

In the Sondheimer oscillation we have pointed out that the predicted
oscillation amplitude does not match with the experimental data
because of the Shubnikov-de Haas oscillation which occurs above $4$ T. In order
to confirm irrelevance of the formation of Landau levels at low
magnetic fields, we take into account the vector potential for the
MR and Hall resistance. This has been performed in the context of
the quantum Hall effect in graphene, where the two valley
contributions are simply added \cite{Graphene_QHE}.

The longitudinal conductance is given by \cite{Graphene_QHE} \bqa
\sigma_{xx}(H, T) &=& \frac{e^{2} N_{f} \Gamma}{4 \pi^{2} T}
\int_{-\infty}^{\infty} d \omega
\frac{1}{\cosh^{2}\Bigl(\frac{\omega+\mu}{2T}\Bigr)}
\frac{\Gamma}{\Bigl(\frac{v_{f}^{2} e H}{c}\Bigr)^{2} + (2 \omega
\Gamma)^{2}} \nn &\times& \Bigl[ 2 \omega^{2} + \frac{(\omega^{2}
+ \Delta^{2} + \Gamma^{2}) \Bigl(\frac{v_{f}^{2} e H}{c}\Bigr)^{2}
- 2 \omega^{2} (\omega^{2} - \Delta^{2} +
\Gamma^{2})\Bigl(\frac{v_{f}^{2} e H}{c}\Bigr)}{(\omega^{2} -
\Delta^{2} - \Gamma^{2})^{2} + 4 \omega^{2} \Gamma^{2}} \nn &-&
\frac{\omega (\omega^{2} - \Delta^{2} + \Gamma^{2})}{\Gamma} \Im
\Psi\Bigl( \frac{\Delta^{2} + \Gamma^{2} - \omega^{2} - 2 i \omega
\Gamma}{2 v_{f}^{2} |e H|/c} \Bigr) \Bigr] \eqa with $N_{f} = 1$
($N_{f} = 2$ for graphene), where $\Delta = \hbar \omega_{L}$ is
the Zeeman energy, $\Gamma$ is the imaginary part of the electron
self-energy due to disorder, and $\Psi(z)$ is the digamma
function.

The Hall conductance is \cite{Graphene_QHE} \bqa &&
\sigma_{xy}(H,T) = \frac{e^{2} N_{f}}{2\pi} \nu_{B} , \eqa where
$\nu_{B}$ is the filling factor, given by \bqa \nu_B &=&
\int_{-\infty}^{\infty} \frac{d \omega}{2\pi} \tanh\Bigl(
\frac{\omega+\mu}{2T} \Bigr) \Bigl[ \frac{\Gamma}{(\omega -
\Delta)^{2} + \Gamma^{2}} + (\omega \longleftrightarrow - \omega)
\nn &+& 2 \sum_{n = 1}^{\infty} \Bigl( \frac{\Gamma}{(\omega -
M_{n})^{2} + \Gamma^{2}} + (\omega \longleftrightarrow - \omega)
\Bigr) \Bigr] , \eqa where $M_{n} = \sqrt{\Delta^{2} + 2 n
v_{f}^{2} |e H|/c}$ is the dispersion of surface Dirac electrons
in the presence of the Landau level.

By using these expressions, we plot the Hall conductance in Fig. 7. First of all,
the plateau in the Hall conductance is clearly shown.
This behavior is far from that given by the experimental data. Therefore,
we conclude that our regime is far from being described by the quantum Hall effect.
The introduction of the Zeeman term is sufficient to explain the
transport data at low magnetic fields.

\begin{figure}[t]
\includegraphics[width=0.4\textwidth]{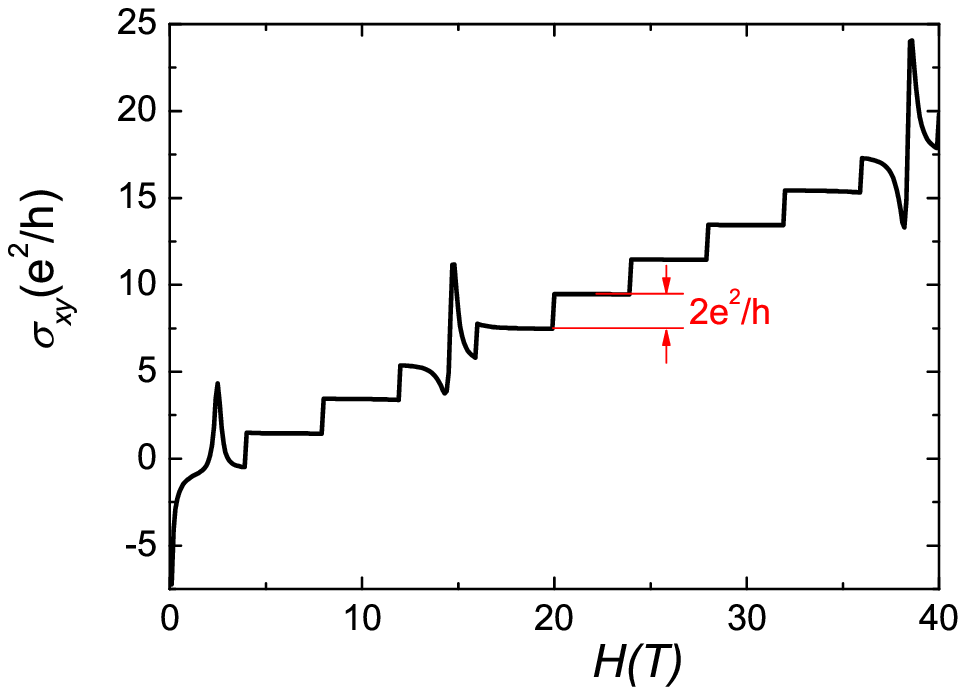}
\caption{Quantized Hall conductance, expected to be relevant in
high magnetic fields.} \label{fig9}
\end{figure}

$^{\ast}$hjkim76@daegu.ac.kr; $^{\dagger}$sasaki@sci.kj.yamagata-u.ac.jp\\